\documentclass[10pt]{article}

\usepackage{amsmath}

\usepackage{array}

\usepackage{appendix}

\usepackage{tocloft}      

\usepackage{graphicx}

\usepackage{amsfonts}

\usepackage{amssymb}

\usepackage{mathrsfs}

\usepackage{yfonts}

\usepackage{euscript}

\usepackage{upgreek}

\usepackage{mathtools}

\usepackage{slantsc}
\usepackage{calligra}

\usepackage{bbold}          

\usepackage[T1]{fontenc}

\usepackage{epsf}

\usepackage{latexsym}

\usepackage{tipa}

\usepackage{makeidx}

\makeindex

\def\be{\begin{equation}}

\def\ee{\end{equation}}

\def\beq{\begin{equation}}

\def\eeq{\end{equation}}

\def\bea{\begin{eqnarray}}

\def\eea{\end{eqnarray}}

\def\ni{\noindent}

\def\foo{\footnote}

\def\!{\hspace{-1.6667em}}

\def\biP{\mbox{\boldmath$P$}}

\def\biQ{\mbox{\boldmath$Q$}}

\def\ma{\mbox{a}}

\def\md{\mbox{d}} 

\def\me{\mbox{e}}

\def\mg{\mbox{g}}

\def\mi{\mbox{i}}

\def\ml{\mbox{l}}

\def\mn{\mbox{n}}

\def\muu{\mbox{u}}

\def\bupSigma{\mbox{\boldmath$\Sigma$}}                 

\def\sa{\mbox{\scriptsize a}}

\def\sf{\mbox{\scriptsize f}}

\def\si{\mbox{\scriptsize i}}

\def\sll{\mbox{\scriptsize l}}  

\def\sn{\mbox{\scriptsize n}}

\def\sS{\mbox{\scriptsize S}}

\def\sT{\mbox{\scriptsize T}}

\def\sV{\mbox{\scriptsize V}}

\def\sfh{\mbox{\sffamily{\scriptsize h}}}     

\def\sfk{\mbox{\sffamily{\scriptsize k}}}     

\def\sfo{\mbox{\sffamily{\scriptsize o}}}     

\def\sfp{\mbox{\sffamily{\scriptsize p}}}     

\def\sfq{\mbox{\sffamily{\scriptsize q}}}     

\def\sfr{\mbox{\sffamily{\scriptsize r}}}     

\def\sfs{\mbox{\sffamily{\scriptsize s}}}     

\def\sfA{\mbox{\sffamily{\scriptsize A}}}      

\def\sfB{\mbox{\sffamily{\scriptsize B}}}      

\def\sfC{\mbox{\sffamily{\scriptsize C}}}      

\def\sfD{\mbox{\sffamily{\scriptsize D}}}      

\def\sfF{\mbox{\sffamily{\scriptsize F}}}      

\def\sfG{\mbox{\sffamily{\scriptsize G}}}      

\def\sfK{\mbox{\sffamily{\scriptsize K}}}      

\def\sfN{\mbox{\sffamily{\scriptsize N}}}      

\def\sfO{\mbox{\sffamily{\scriptsize O}}}      

\def\sfU{\mbox{\sffamily{\scriptsize U}}}      

\def\sfV{\mbox{\sffamily{\scriptsize V}}}      

\def\sfW{\mbox{\sffamily{\scriptsize W}}}      

\def\sfZ{\mbox{\sffamily{\scriptsize Z}}}      

\def\scC{\mbox{\scriptsize ${\cal C}$}}          

\def\scH{\mbox{\scriptsize ${\cal H}$}}          

\def\scO{\mbox{\scriptsize ${\cal O}$}}

\def\FrO{\mbox{$\mathfrak{O}$}}                   

\def\FrL{\mbox{$\mathfrak{L}$}}                   
					  												  
\def\bFrb{\mbox{\Large $\mathfrak{b}$}}           
 
\def\FrC{\mbox{$\mathfrak{C}$}}                   

\def\FrD{\mbox{$\mathfrak{D}$}}	                  
\def\lFrg{\mbox{\Large \textfrak{g}}}             
\def\FrH{\mbox{$\mathfrak{H}$}}                   
\def\bFrj{\mbox{\Large $\mathfrak{j}$}}           

\def\bFrM{\mbox{\boldmath$\mathfrak{M}$}}         

\def\FrN{\mathfrak{N}}                            
                                                  
\def\FrS{\mbox{\Large $\mathfrak{s}$}}            
\def\FrT{\mathfrak{T}}                            
              
\def\FrU{\mbox{$\mathfrak{U}$}}                   
\def\FrX{\mathfrak{X}}                            
\def\Frg{\mbox{\normalsize $\mathfrak{g}$}}                                  

\def\Frk{\mbox{\scriptsize $\mathfrak{K}$}}                                  

\def\Frh{\mbox{$\mathfrak{h}$}}

\def\Frg{\mbox{$\mathfrak{g}$}} 

\def\Frh{\mbox{$\mathfrak{h}$}}                 

\def\Frk{\mbox{$\mathfrak{k}$}}

\def\Fru{\mbox{$\mathfrak{u}$}}

\def\K{Kucha\v{r} }

\def\5Star{\mbox{\Large$\star$}}              

\def\cr{\mbox{\scriptsize{\bf $\mbox{ } \times \mbox{ }$}}}

\def\sumi2{\sum\mbox{}_{\mbox{}_{\mbox{\scriptsize $i$=1}}}^2}

\def\sumi3{\sum\mbox{}_{\mbox{}_{\mbox{\scriptsize $i$=1}}}^3}

\def\sumABcycles3{\sum\mbox{}_{\mbox{}_{\mbox{\scriptsize cycles  $A,B$=1}}}^{3}}

\def\sumCDcycles3{\sum\mbox{}_{\mbox{}_{\mbox{\scriptsize cycles  $C,D$=1}}}^{3}}

\def\sumj3{\sum\mbox{}_{\mbox{}_{\mbox{\scriptsize $j$=1}}}^3}

\def\sumk3{\sum\mbox{}_{\mbox{}_{\mbox{\scriptsize $k$=1}}}^3}

\begin{document}


\begin{center}

{\Large{\bf ON TYPES OF OBSERVABLES IN CONSTRAINED THEORIES}}

\vspace{.1in}

{\bf Edward Anderson} 

\vspace{.1in}

{\em DAMTP, Centre for Mathematical Sciences, Wilberforce Road, Cambridge CB3 OWA.} \normalsize

\end{center}

\vspace{.1in}

\begin{abstract}

The \K observables notion is shown to apply only to a limited range of theories. 
Relational mechanics, slightly inhomogeneous cosmology and supergravity are used as examples that require further notions of observables.  
A suitably general notion of A-observables is then given to cover all of these cases. 
`A' here stands for `algebraic substructure'; A-observables can be defined by association with each closed algebraic substructure of a theory's constraints.  
Both constrained algebraic structures and associated notions of A-observables form bounded lattices. 

\end{abstract}


\section{Introduction} 

Consider a classical constrained theory \cite{Dirac, HT92, ABook, ABrackets}, 
with configurations $Q^{\sfA}$, 
conjugate momenta   $P_{\sfA}$, 
classical brackets  $\mbox{\bf |[ } \mbox{ } \mbox{\bf ,} \mbox{ } \mbox{\bf ]|}$ 
and constraints     ${\cal C}_{\sfC}$.
{\it Observables} \cite{DiracObs, Dirac, HT92, Kuchar92, I93, Kuchar93, ABeables} are then objects forming zero brackets with the constraints, 
\beq
\mbox{\bf |[} {\cal C}_{\sfC} \mbox{\bf ,} \,  O_{\sfO} \mbox{\bf ]|} \mbox{ `=' } 0 \mbox{ } .
\label{Basic-Obs}
\eeq
There are multiple such notions due to `zero', `brackets' and `the constraints' taking a variety of different precise meanings \cite{ABeables}.
Some such notions of observables can be more physically useful than just any functions (or functionals) of $Q^{\sfA}$ and $P_{\sfA}$.
This is because of their greater physical content; moreover, at least some versions of such notions of observables would be expected to contain physical information only.  
At the very least this property is required in phrasing final answers to physical questions about a theory.


\mbox{ }

\ni The zero can for instance be the usual notion of zero -- alias  {\it strongly zero} in this context --
                             or Dirac's notion  \cite{Dirac} of {\it weakly zero}, denoted $\approx$, meaning zero up to terms linear in constraints.
The bracket can for instance, at the classical level, be the usual Poisson bracket; we restrict attention to this case in the rest of this Article.
The constraints can include all the constraints, or a subset thereof.
Three particular cases of this which are well-known enter the current article.\footnote{See e.g. \cite{Bergmann61, BK72, PSS10} for the further distinct Bergmann concept of observables.}

\mbox{ }

\ni 1) {\it Classical Dirac observables} \cite{DiracObs, Dirac, HT92} are quantities $D_{\sfD}$ that 
(usually weakly Poisson) brackets-commute with {\sl all} of a given theory's first-class constraints, 
\beq
\mbox{\bf |[} {\cal C}_{\sfC} \mbox{\bf ,} \, D_{\sfD} \mbox{\bf ]|} \approx 0 \mbox{ } .
\label{C-D}
\eeq 
Various interpretations proposed for these entities gave further colourful names for these, 
such as `evolving constants of the motion' \cite{Rov91a} and `perennials' \cite{Kuchar93, Kuchar99, BF08, Kouletsis08}.
The names `true' and `complete' observables have also been used in similar contexts \cite{Rov91a, Rov02b, Thiemann}.
Conceptually, these are `{\it maximally constrained}' observables; 
the current Article will also give mathematical cause to call these `{\it zeroth observables}'.

\mbox{ }

\ni 2) `{\it Unconstrained observables}' $U_{\sfU}$, on the other hand have no constraint-related restrictions.
`{\it Unital observables}' is also a mathematical name for these for reasons which will become clear as this article develops.
`Partial observables' \cite{RovelliBook, Ditt, Dittrich, Tambornino} are a well-known particular notion of unconstrained observables 
(though these play no further part in the current article).

\mbox{ }

\ni 3) \K introduced \cite{Kuchar93} a further notion of observables, which subsequently indeed came to be termed {\it \K observables}.   
These are quantities $K_{\sfK}$ which (again usually weakly Poisson) brackets-commute with all of a given theory's first-class linear constraints, 
\beq
\mbox{\bf |[}    {\cal F}\ml\mi\mn_{\sfN}    \mbox{\bf ,}  \, K_{\sfK}    \mbox{\bf ]|} \approx 0 \mbox{ } .
\label{FLIN-K}
\eeq
\ni Using \K observables reflects treating first-class quadratic constraints, ${\cal Q}\muu\ma\md$, 
                           distinctly from first-class linear ones            ${\cal F}\ml\mi\mn_{\sfN}$ \cite{Kuchar92, Kuchar93, Kuchar99, Kouletsis08, BF08}. 
%
%
The Problem of Time literature \cite{Kuchar92, I93, BF08, APoT, APoT2, FileR, APoT3, ABook} is replete with reasons why this may be desirable; 
moreover also \K observables are more straightforward to find than Dirac ones.
Indeed, the {\it Problem of Observables}  -- that it is hard to construct a sufficiently large set of these to describe all physical quantities, especially for Gravitational Theory -- 
is one of the many facets of the Problem of Time.
\K observables are more straightforward to construct than Dirac ones, and classical ones are more straightforward to construct than quantum ones.
The Problem of Observables remains little understood.
Strategies proposed in this regard include the following.

\mbox{ } 

\ni Strategy 1) Insist upon constructing Dirac observables, $D_{\sfD}$.

\mbox{ } 

\ni Strategy 2) Make use of unconstrained observables, $U_{\sfU}$ \cite{RovelliBook, Ditt, Dittrich, Tambornino}.

\mbox{ } 

\ni In Strategy 2) case one is disregarding all the information content of the constraints.  
However, on some occasions Strategy 2) is used as a stepping stone to constructing Dirac-type observables.

\mbox{ } 

\ni Strategy 3) Consider \K observables, $K_{\sfK}$, to suffice.  

\mbox{ } 

\ni Some examples are also useful at this stage. 
In unconstrained theories, all three notions coincide.
In theories whose sole constraint is first-class quadratic, unconstrained and \K observables are coincident, trivial and distinct from Dirac observables, which are nontrivial.
This applies to minisuperspace models and to some of the simplest relational mechanics \cite{FileR, AConfig} models.
In Electromagnetism and Yang--Mills Theory, on the other hand, all the constraints are first-class linear, 
so Dirac and \K observables coincide and are distinct from unconstrained observables.
Finally, in more advanced relational mechanics theories \cite{AMech} and in GR, there are first-class linear constraints and a first-class quadratic constraint. 
Here, none of the above three notions of observables coincide.
In particular, for GR the first-class linear constraints are the vector-valued GR momentum constraint ${\cal M}_i$, 
whereas the first-class quadratic constraint is the scalar-valued GR Hamiltonian constraint ${\cal H}$.
The \K observables then obey 
\beq
\mbox{\bf \{}    {\cal M}_{i}    \mbox{\bf ,}  \, K_{\sfK}    \mbox{\bf \}} \approx 0 \mbox{ } , 
\label{K_GR}
\eeq
where $\mbox{\bf \{}  \mbox{ } \mbox{\bf ,}  \mbox{ }   \mbox{\bf \}}$ is specifically the Poisson bracket.   
For relational mechanics, ${\cal E}$ which has the mathematical form of a quadratic energy constraint takes over the role of ${\cal H}$. 
Here many examples' \K observables are known explicitly \cite{ABeables2}: they are shapes, or shapes and scales, for various different notions of shape \cite{AMech}, 
and conjugate quantities.
For Electromagnetism, the \K observables' forms are also well-understood \cite{ABeables}, however for GR they remain only formally understood \cite{ABeables2}.

{            \begin{figure}[ht]
\centering
\includegraphics[width=0.4\textwidth]{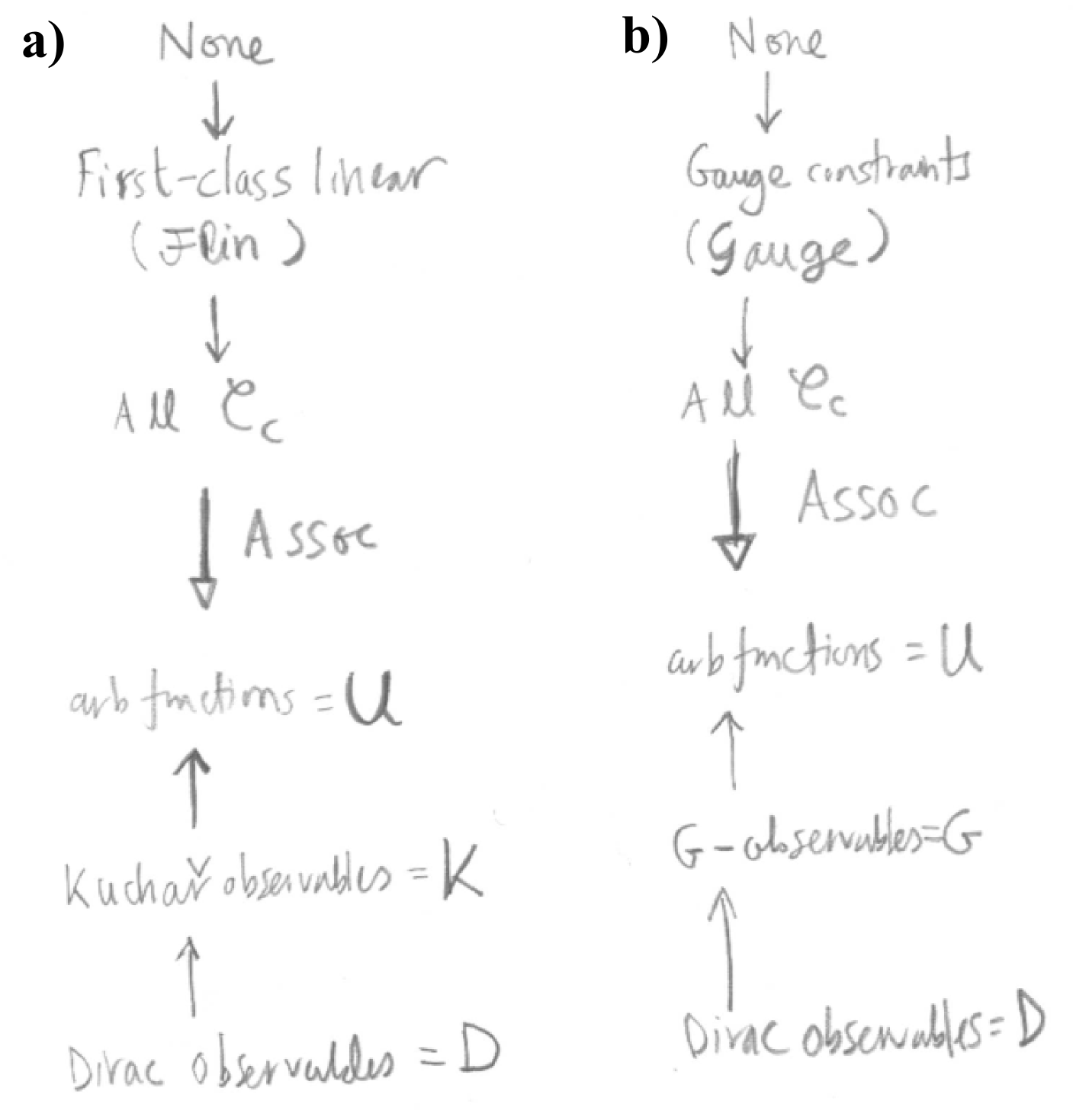} 
\caption[Text der im Bilderverzeichnis auftaucht]{        \footnotesize{a) The incipient position on constraint algebraic structures. 
b) With Dirac's conjecture proving to be false, more thought in general needs to be put in as regards whether the intermediate algebraic structure is first-class linear or gauge.
The arrows in the upper half indicate inclusion of further generators, in ways allowed by the integrability relations between these.
The arrows in the lower half indicate inclusion of further relations.
$Assoc$ is the map by which notions of observables are associated to sets of constraints.} }
\label{Latt-1} \end{figure}          }

\ni The next consideration of note is Dirac's conjecture, by which first-class linear constraints would always coincide with gauge constraints.
Unfortunately, this is false \cite{HT92, BF08}.
Due to this, a fourth notion of observable is unearthed: the $\lFrg$-observables $G_{\sfZ}$ which commute with all of a theory's gauge constraints: 
\beq
\mbox{\bf |[} G_{\sfZ} \mbox{\bf ,} \, {\cal G}\ma\muu\mg\me_{\sfG} \mbox{\bf ]|} \approx 0 \mbox{ } ;
\eeq
$\lFrg$ is that Gauge Theory's underlying gauge group.
Of course, for some theories ${\cal F}\ml\mi\mn_{\sfN}$ and ${\cal G}\ma\muu\mg\me_{\sfG}$ coincide, in which case $K_{\sfK}$ and $G_{\sfZ}$ coincide as well; 
e.g. this is the case in relational mechanics, Electromagnetism and Yang--Mills Theory.  

\mbox{ } 

\ni Strategy 4) is then to consider $\lFrg$-observables observables to suffice.

\mbox{ } 

\ni The next development follows from perceiving that  a notion of observables being meaningful 
is contingent on the subset of constraints it commutes with a fortiori closing as a subalgebraic structure \cite{ABeables} (Lemma 1 of Appendix \ref{Observables}).
In particular, in some cases the ${\cal F}\ml\mi\mn_{\sfN}$ do not close, and thus indeed the notion of \K observables $K_{\sfK}$ does not exist for that theory.
In Problem of Time considerations, this means that many approaches to the Problem of Observables can only be addressed once the Constraint Closure Problem has been resolved \cite{ABook}. 
I.e. one needs the constraints algebraic structure prior to consideration of associated observables algebraic structures.
It is also required that the notion of bracket involved in defining the $O_{\sfO}$ matches that under which the ${\cal C}_{\sfC}$ close.
Additionally, by Appendix \ref{Observables}'s Lemma 2, the $O_{\sfO}$ themselves form a closed algebraic structure.
[It is, moreover, interesting that observables algebraic structures exhibit some conceptual similarities with theory of Casimirs: Appendix \ref{Associated}, 
which proved to be a central ingredient in understanding (much simpler) algebraic structures: Lie groups and their reps.]  

\mbox{ } 

\ni Along the above lines, focus shifts from \K observables to the observables associated with whichever closed subalgebraic structures 
each theory's constraint algebraic structure happens to possess.
I term these A-observables, where the `A' stands for `algebraic substructure'.
The subalgebraic structures of a gives algebraic structure then turn out to form a bounded lattice (Appendix \ref{Lattices}) under the inclusion operation.
The association by `forms zero brackets with' then produces in a further lattice, now of notions of observables.
This pair of bounded lattice structures form the current Article's main underlying result.
Fig \ref{Latt-1} gives preliminary versions of this involving $U_{\sfU}$, $D_{\sfD}$, $K_{\sfK}$ and $G_{\sfZ}$.
The current Article illustrates this with nontrivial examples of notions of A-observable arising in relational mechanics, slightly inhomogeneous cosmology and Supergravity, 
in Secs 2, 3 and 4 respectively.  
We conclude in Sec 4 with a more general account of the lattice of constraint subalgebraic structures and of the associated lattice of notions of A-observables.  
Discussion of the Article's range of algebraic concepts is difered to a multi-part Appendix.

\section{Relational mechanics examples}

The constraint algebraic structures for these are outlined Appendix \ref{CAS}. 
Some of the constraint subalgebras are then given in Fig \ref{Latt-2}, alongside the associated notions of A-observables. 

{            \begin{figure}[ht]
\centering
\includegraphics[width=1.0\textwidth]{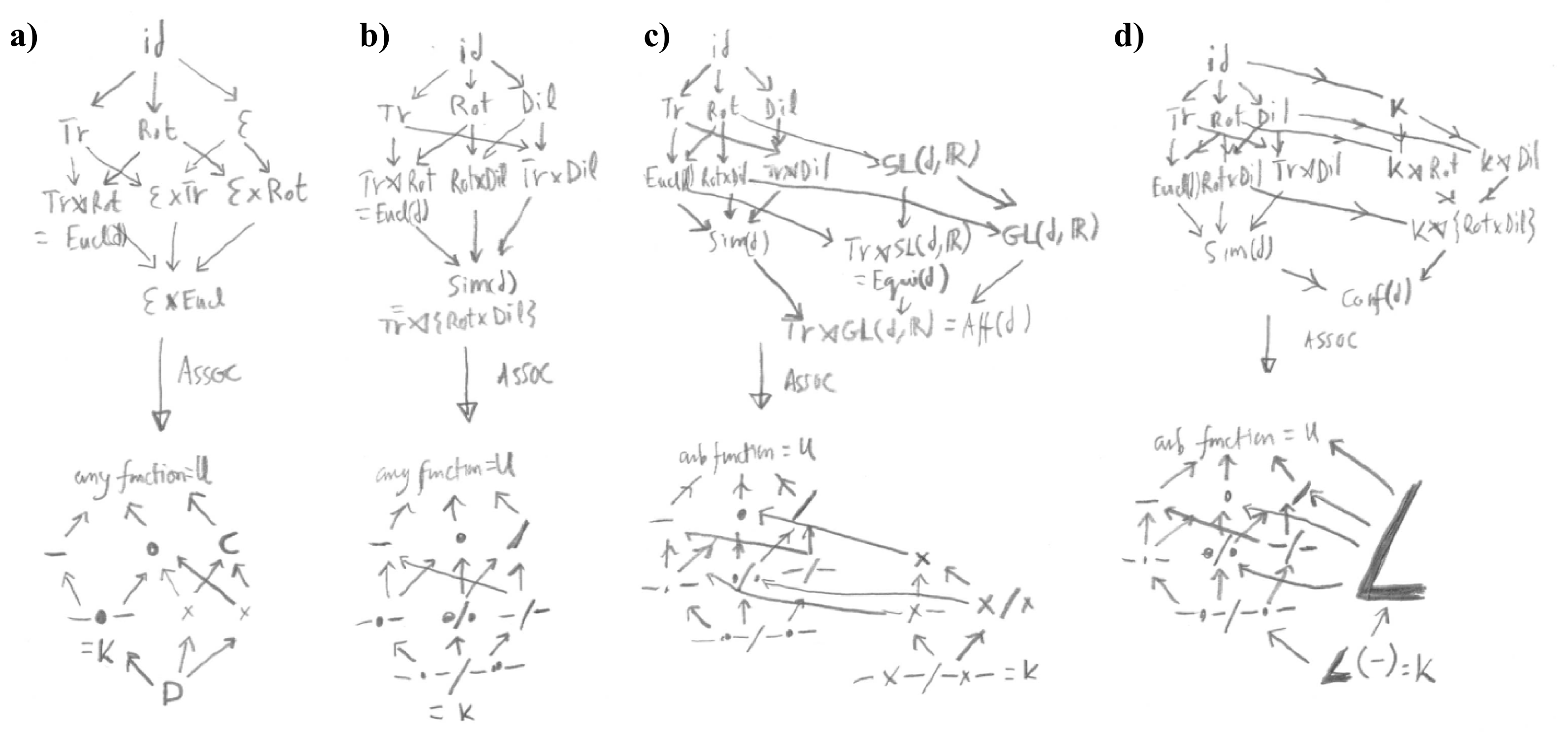} 
\caption[Text der im Bilderverzeichnis auftaucht]{        \footnotesize{Constraint subalgebraic structures for a range of relational mechanics models, 
with corresponding notions of A-observables. 
a) is for Euclidean relational mechanics, including the quadratic energy constraint ${\cal E}$, which forms its own separate subalgebraic structure, as per Appendix \ref{CAS}.
I term the corresponding notion of observables {\it chronos observables} due to the ties between ${\cal E}$ and time provision in whole-universe models \cite{APoT3, ABook}.
Because ${\cal E}$ combines in the same trivial manner in each of the subsequent examples, it is dropped from the presentation of the rest of these.
b) is similarity relational mechanics, c) is affine relational mechanics and d) is conformal relational mechanics; consult Appendix \ref{CAS} for the 
nomenclature used, and \cite{AMech} for a more detailed presentation.  
- denotes differences (relative particle separation vectors, or possibly particle cluster separation ones), $\cdot$ denotes dot product quantities, $\/$ denotes ratios; 
these furthermore combine when multiple constraints apply, as indicated.
Upon passing to the affine case, cross products $\cr$ supplant dot products, whereas upon passing to the conformal case, local angle quantities $\angle$ appear.  
Note in this case that the lattice morphism is not 1 to 1, since angles are already ratios and combinations of dot products. } }
\label{Latt-2} \end{figure}          }

\section{Slightly inhomogeneous cosmology example}

In this model \cite{HallHaw, SIC, SIC-2} $\bupSigma = \mathbb{S}^3$, and one splits modewise 
(each mode corresponds to collection of $\mathbb{S}^3$-harmonics labels) and into scalar, vector and tensor (SVT) parts.  
This splits ${\cal H}$   into $\mbox{}^{\sS}{\cal H}$,    $\mbox{}^{\sV}{\cal H}$ and $\mbox{}^{\sT}{\cal H}$ pieces, 
and         ${\cal M}_i$ into $\mbox{}^{\sS}{\cal M}$ and $\mbox{}^{\sV}{\cal M}$                             pieces.
Modewise and SVT-wise splits give decoupled equations at the unreduced level, to lowest nontrivial perturbative order.  

\mbox{ } 

\ni It then turns out that \cite{SIC, SIC-2} taking out $\mbox{}^{\sS}{\cal M}$ and $\mbox{}^{\sV}{\cal M}$ fails to take out all of the $Diff(\mathbb{S}^3)$ information left in the model.

\ni The amount by which it fails to do so depends on whether one is considering the single minimally coupled scalar field matter version \cite{HallHaw, SIC} 
                                                                                                                   or the vacuum version \cite{SIC-2}.
In the vacuum case, it is out by 1 configuration space degree of freedom, whereas it is out by 2 in the scalar field case.
The vacuum case also has a more decoupled partly reduced system, 
by which it is clear \cite{SIC-2} that removing $\mbox{}^{\sV}{\cal H}$ suffices to take one down to $Diff(\mathbb{S}^3)$ invariance, 
i.e. to this model's counterpart of $Superspace(\mathbb{S}^3) = \mbox{Riem}(\mathbb{S}^3)/Diff(\mathbb{S}^3)$.
It is this vacuum case that the current Article considers in detail; the scalar field counterpart remains only partly understood.
In this vacuum case, then, the constraint algebraic structure is as in Fig \ref{Latt-3}.a) and the corresponding notions of observables are as in Fig \ref{Latt-3}.b).

{            \begin{figure}[ht]
\centering
\includegraphics[width=0.6\textwidth]{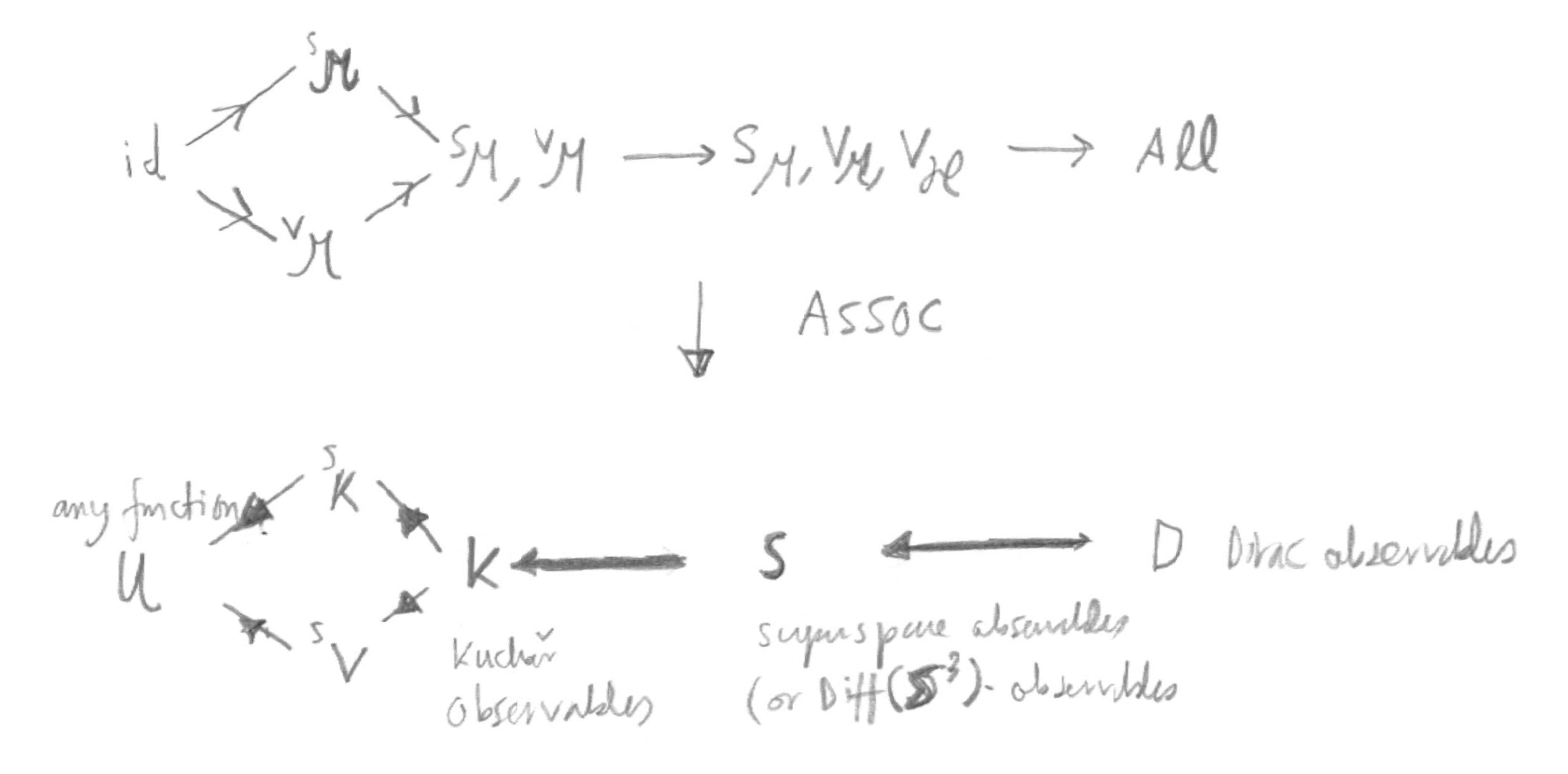} 
\caption[Text der im Bilderverzeichnis auftaucht]{        \footnotesize{A range of interesting notions of constraint subalgebraic structure and of A-observables for 
slightly inhomogeneous cosmology.} }
\label{Latt-3} \end{figure}          }

\ni Firstly, note that \K observables and $\lFrg$-observables -- for $\lFrg = Diff(\mathbb{S}^3)$ -- are distinct in this example.
I also term the latter Superspace observables.  

\mbox{ }  

\ni Secondly, note that there are additional notions of A-observables, for which I suggest the names S-\K observables and V-\K observables, 
meaning commutation with just $\mbox{}^{\sS}{\cal M}$ and just $\mbox{}^{\sV}{\cal M}$ respectively.   
Taking out just $\mbox{}^{\sV}{\cal M}$ has a scalar field case parallel which underlies Wada's even more partial reduction in \cite{Wada85}; 
he then proceeded to reduce out $\mbox{}^{\sS}{\cal M}$ as well \cite{Wada}.

\section{Supergravity example} 

Introducing a local Lorentz frame formulation to accommodate the model's fermionic species 
causes a local Lorentz frame constraint ${\cal J}$ (and its conjugate) to appear.\footnote{In the current Article's schematic presentation, I drop the spinorial indices. 
See e.g. \cite{DEath, VM10} for more details.}

\mbox{ } 

\ni The canonical formulation of Supergravity then has the expected ${\cal H}$, ${\cal M}_i$ and ${\cal J}$ constraints alongside a specifically supersymmetric constraint ${\cal S}$.  
All we need to know for this Article about ${\cal J}$ and ${\cal S}$ are that each of these constraints is linear in the momenta, 
and the schematic form of the subsequent constraint brackets, as per Appendix \ref{CAS}.
The particular feature this Article concentrates upon is that in Supergravity a subset of the linear constraints -- the supersymmetry constraints -- 
have the Supergravity counterpart of the quadratic $\scH$ as {\sl their} integrability; this follows from eq (\ref{S-S->H}).
This means that \K observables are not well-defined for Supergravity, since the quadratic ${\cal H}$ now arises as an integrability of the linear ${\cal S}$.  
Eq (\ref{S-S->H}) forms the second integrability of the `two-way' pair, to the first integrability being of form (\ref{Dirac-Algebroid}).

\mbox{ } 

\ni Some constraint subalgebraic structures of note are then provided in Fig \ref{Latt-3}.  
In particular, the non-supersymmetric linear constraint combination of ${\cal J}$ and ${\cal M}_i$ closes, as does the GR-like combination of ${\cal H}$ and ${\cal M}_i$, 
and the non-supersymmetric combination of  ${\cal J}$, ${\cal M}_i$ and ${\cal H}$.
I term the corresponding notions of A-observables non-supersymmetric \K, GR-likes and non-supersymmetric Dirac \cite{AMech}.  
%
{            \begin{figure}[ht]
\centering
\includegraphics[width=0.5\textwidth]{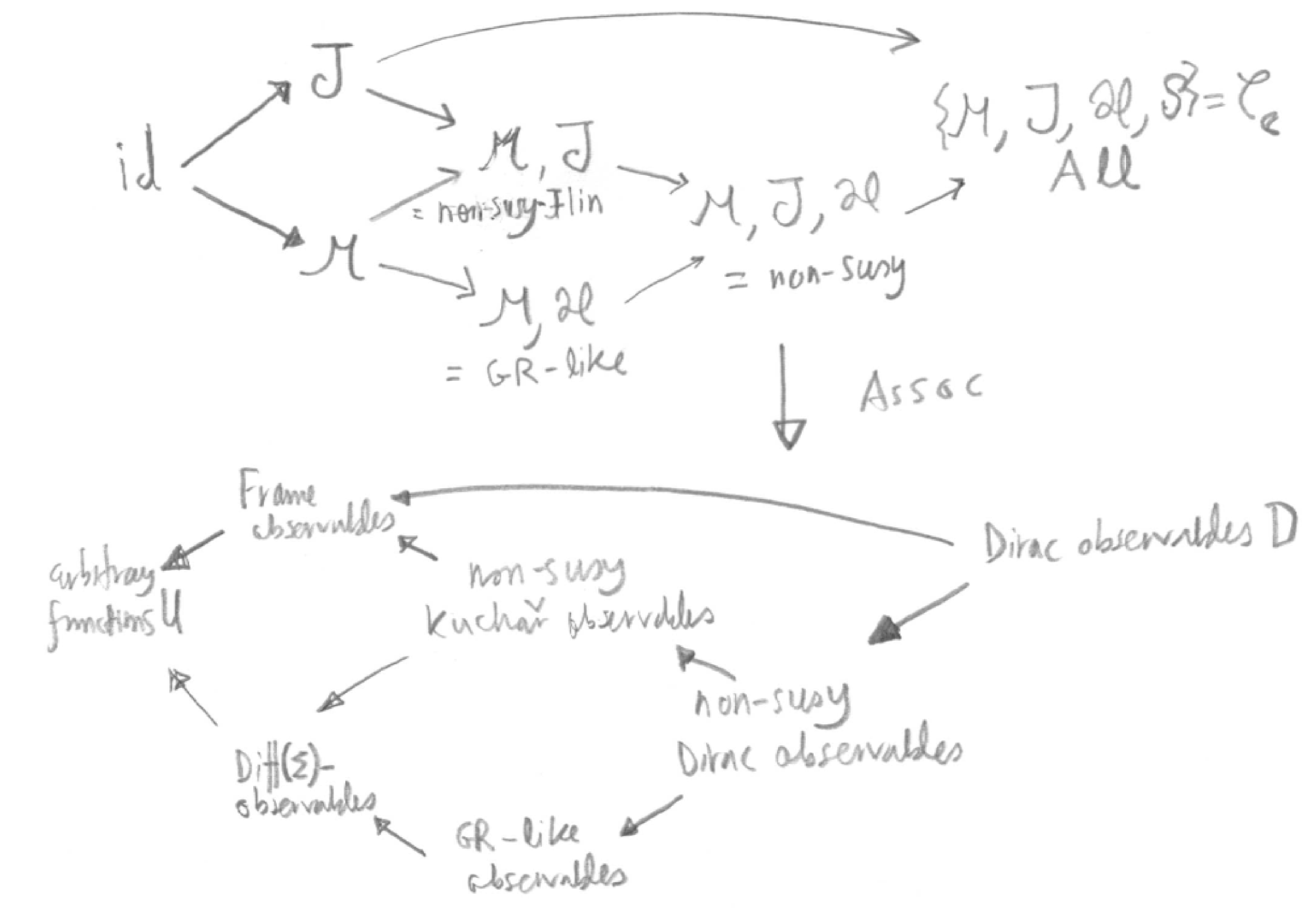} 
\caption[Text der im Bilderverzeichnis auftaucht]{        \footnotesize{A range of interesting notions of constraint subalgebraic structure and of A-observables for Supergravity.} }
\label{Latt-4} \end{figure}          }

\vspace{3in}

\section{Conclusion: lattice of A-observables notions}

There is a notion of observables associated with each constraint subalgebraic structure that a given theory's constraint algebraic structure happens to possess.
I term these, collectively, A-observables (Fig \ref{Latt-5}).
Both the constraint subalgebraic structures of a given theory's constraint algebraic structure, and the theory's notion of A-observables form bounded lattice structures.
The Dirac observables $D_{\sfD}$ correspond to the zero of the latter lattice, 
arising from considering the whole constraint algebraic structure, which is the unit of the former lattice.
On the other hand, the unconstrained observables $U_{\sfU}$ correspond to the unit of the latter lattice, 
arising from considering the trivial subalgebraic structure (id), which is the zero of the former lattice.
Once this context has been framed, it is clear that the notion \K observables -- commuting with first-class linear constraints -- is only one some occasions well-defined.  
This becomes seen, rather, as an early example of nontrivial A-observable (i.e. neither unconstrained nor Dirac), 
which is realized in some theories -- such as GR -- but not others -- such as Supergravity.
Thus approaches to Quantum Gravity which depend on the Quadratic to linear distinction between constraints, and the subsequent notion of \K observables, can be fragile to change of theory. 
This is to be contrasted with how $D_{\sfD}$ and $U_{\sfU}$ are defined for {\sl all} theories.  

{            \begin{figure}[ht]
\centering
\includegraphics[width=0.5\textwidth]{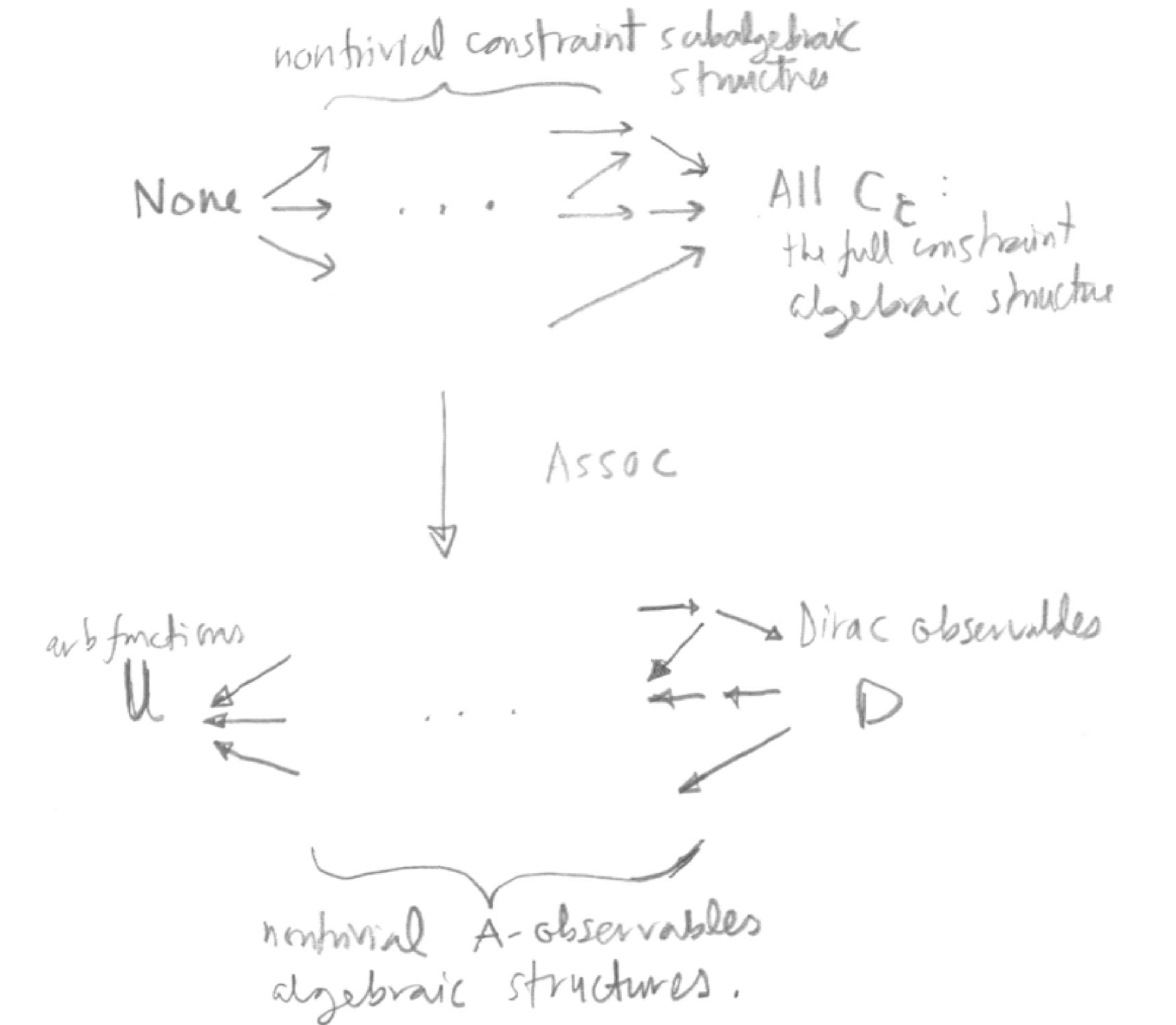} 
\caption[Text der im Bilderverzeichnis auftaucht]{        \footnotesize{The logical conclusion of the small linear ordering of notions of observables 
in Fig 1.a) is the lattice of A-observables.} }
\label{Latt-5} \end{figure}          }

\ni Moreover, other interesting closed subalgebraic structures arise, each with an associated notion of A-observable.
In the current Article, we have discussed a number of such, and coined names for them.

\mbox{ } 

\ni A) Frame observables, non-supersymmetric \K observables, GR-like observables and non-supersymmetric Dirac observables in the case of Supergravity. 

\ni B) Superspace observables as distinct from \K observables in slightly inhomogeneous cosmology, alongside scalar-only and vector-only \K observables.   

\ni C) A wide range of geometrically meaningful A-observables in the context of relational theories of mechanics.  

\mbox{ } 

\ni The Introduction's list of strategies toward dealing with the Problem of Observables can then be updated to include the following.

\mbox{ } 

\ni Strategy 5) Conceive in terms of the lattice of notions of A-observables $A_{\sfO}$, such that 
\beq
\mbox{\bf |[} A_{\sfO} \mbox{\bf ,} \, {\cal C}_{\sfW} \mbox{\bf ]|} `=' 0 \mbox{ } ,
\eeq 
for ${\cal C}_{\sfW}$ whichever (closed) subalgebraic structure of the constraint algebraic structure.
This strategy covers 

\mbox{ } 

\ni 1) making whichever such choice. 
              
\ni 2) The possibility of having further selection principles among the various candidate theories 
whose constraint algebraic structures admit a variety of nontrivial proper subalgebraic structures.

\ni 3) Using one or more notions of A-observables as intermediate stages in computing more restrictive types of observables.  

\mbox{ }

\ni Note moreover that the Dirac observables strategy is both {\sl always available} and {\sl requires no choice or further selection principle}.  
This is a direct consequence of taking the whole constraint algebraic structure, as follows.  
The same two maxims also apply at the opposite extreme of the lattice of subalgebraic structures: the unconstrained observables.  

\mbox{ } 

\ni I end by pointing out two future research directions.

\mbox{ } 

\ni 1) Increase the scope of the examples used to include more supersymmetric relational mechanics as well as superconformal supergravity.

\ni 2) Quantum mechanical counterparts and quantization process from classical to quantum versions remain to be presented \cite{ABrackets}.

\mbox{ }  

\noindent {\bf Acknowledgements} 
I thank those close to me.
I thank the fqXi and DAMTP for travel money and hosting in 2014-2015 (slightly inhomogeneous cosmology) and 2015-2016 (Supergravity).
I also thank Jeremy Butterfield, Chris Isham, Malcolm MacCallum, Don Page, Enrique Alvarez, John Barrow, Marc Lachieze-Rey and Reza Tavakol for discussions or helping me with my career.  

\begin{appendices}

\section{Algebraic structures}\label{Appendix}

\subsection{Lie groups and Lie algebras}\label{Lie}

\ni {\it Lie groups} $\lFrg$ \cite{Gilmore} are simultaneously groups and differentiable manifolds; 
additionally their composition and inverse operations are differentiable.  
Working with the corresponding infinitesimal `tangent space' around $\lFrg$'s identity element  -- the Lie algebra $\Frg$ -- 
is more straightforward due to vector spaces' tractability, while very little information is lost in doing so.
For instance, the representations of $\Frg$ determine those of $\lFrg$.  

\mbox{ } 

\ni More formally, a {\it Lie algebra} is a vector space equipped with a product (bilinear map) 
$\mbox{\bf |[} \mbox{ } \mbox{\bf ,} \mbox{ } \mbox{\bf ]|}: \Frg \times \Frg \longrightarrow \mbox{\Frg}$ that is antisymmetric 
and obeys the Leibniz (product) rule and the {\it Jacobi identity}
\beq
\mbox{\bf |[} g_1 \mbox{\bf ,} \, \mbox{\bf |[} g_2 \mbox{\bf ,} \, g_3 \mbox{\bf ]|} \, \mbox{\bf ]|} + \mbox{cycles} = 0 
\label{Jacobi-id}
\ee
$\forall \, g_1, g_2, g_3 \, \in \, \Frg$.
This an example of {\it algebraic structure}: equipping a set with one or more further product operations.  
Particular subcases of Lie brackets include the familiar Poisson brackets and quantum commutators.  

\mbox{ } 

\ni Moreover, a Lie algebra's generators (Lie group generating infinitesimal elements) $\tau_{\sfp}$ obey 
\beq
\mbox{\bf |[} \tau_{\sfp} \mbox{\bf ,} \, \tau_{\sfq} \mbox{\bf ]|} = {C^{\sfr}}_{\sfp\sfq}\tau_{\sfr} \mbox{ } ,
\label{Str-Const}
\eeq 
where ${C^{\sfr}}_{\sfp\sfq}$ are the {\it structure constants} of that Lie algebra.\foo{If functions occur in this role instead, 
then one has strayed into mathematics more complicated than that of Lie algebras; see Appendix \ref{Algebroids} if interested.}
%
It readily follows that 
the structure constants with all indices lowered are totally antisymmetric, and also obey  
\be
{C^{\sfo}}_{[\sfp\sfq}{C^{\sfr}}_{\sfs]\sfo} = 0  \mbox{ } . 
\label{firstJac}
\ee  
%
%
Next suppose that a hypothesis is made about some subset $\Frk$ of the generators, $k_{\sfk}$, being significant.    
These are linear and quadratic constraints in the context of constraints in Gravitational Theory.
Denote the rest of the generators -- which form $\Frh$ -- by $h_{\sfh}$.    
On now needs to check to what extent the algebraic structure in question actually complies with this assignation of significance.    
Such checks place limitations on how generalizable some intuitions and concepts which hold for simple examples of algebraic structures are.  
In general, the split algebraic structure is of the form 
\beq
\mbox{\bf |[} k_{\sfk} \mbox{\bf ,} \,  k_{\sfk^{\prime}} \mbox{\bf ]|} = {C^{\sfk^{\prime\prime}}}_{\sfk\sfk^{\prime}}k_{\sfk^{\prime\prime}} 
                                                                        + {C^{\sfh}}_{\sfk\sfk^{\prime}}               h_{\sfh}                    \mbox{ } ,
\label{Lie-Split-1}
\eeq 
\beq
\mbox{\bf |[} k_{\sfk} \mbox{\bf ,} \,  h_{\sfh}          \mbox{\bf ]|} = {C^{\sfk^{\prime}}}_{\sfk\sfh}k_{\sfk^{\prime}} 
                                                                        + {C^{\sfh^{\prime}}}_{\sfk\sfh}h_{\sfh^{\prime}}                           \mbox{ } ,
\label{Lie-Split-2}
\eeq 
\beq
\mbox{\bf |[} h_{\sfh} \mbox{\bf ,} \,  h_{\sfh^{\prime}} \mbox{\bf ]|} =  {C^{\sfk}}_{\sfh\sfh^{\prime}}               k_{\sfk}                
                                                                        +  {C^{\sfh^{\prime\prime}}}_{\sfh\sfh^{\prime}}h_{\sfh^{\prime\prime}}     \mbox{ } .
																	\label{Lie-Split-3}
\eeq 
Denote the second to fifth right-hand-side terms by 1) to 4).  
1) and 4) being zero are clearly subalgebra closure conditions. 
2) and 3) are `interactions between' $\Frh$ and $\Frk$.
The following cases of this are then realized in this Article.  

\mbox{ } 


\ni I)   {\it Direct product} If 1) to 4) are zero,   then $\Frg = \Frk  \times \Frh$: the {\it direct product group}: 
the Cartesian product of the sets of $\Frk$ and $\Frh$, with group operation $(h_1, k_1) \circ (h_2, k_2) = (h_1 \circ h_2, k_1 \circ k_2)$.  
In this case, one can trivially make use of the Representation Theory of $\Frk$ and $\Frh$ to build that of $\Frk  \times \Frh$.  

\mbox{ } 

\ni II)  {\it Semi-direct product} If 2) alone is nonzero, then $\Frg = \Frk \rtimes \Frh$. 
Now, $\lFrg = \FrN \rtimes \FrH$ is given by $(n_1, h_1) \circ (n_2, h_2) = (n_1 \circ \varphi_{h_1}(n_2), h_1\circ h_2)$ for $\langle \FrN, \circ \rangle  \lhd \lFrg$, 
                                                                                                                              $\langle \FrH, \circ \rangle$ a subgroup of $\lFrg$ 
                                                                                                                          and $\varphi:\FrH \rightarrow \mbox{Aut}(\FrN)$ a group homomorphism.
In this case, {\it Mackey Theory} -- an advanced type of induced representations method -- \cite{Serre, Mackey, I84}
can be used to construct the Representation Theory of semidirect product groups.

\mbox{ } 

\ni III) `{\it Thomas integrability}'. If 1) is nonzero, then $\Frk$ is not a subalgebra: attempting to close it leads to some $k_{\sfk}$ are discovered to be integrabilities.
I denote this by $\Frk \, \mbox{\textcircled{$\rightarrow$}} \, \Frh$.
A simple example of this occurs in splitting the Lorentz group's generators up into rotations and boosts: 
the group-theoretic underpinning \cite{Gilmore} of Thomas precession 
Schematically, decompose the familiar Lorentz algebra bracket into 
\beq
\mbox{\bf |[}J \mbox{\bf ,}  \, J\mbox{\bf ]|} \sim J     \mbox{ } , \mbox{ } \mbox{ } 
\mbox{\bf |[}J \mbox{\bf ,}  \, K\mbox{\bf ]|} \sim K     \mbox{ } , \mbox{ } \mbox{ } 
\mbox{\bf |[}K \mbox{\bf ,}  \, K\mbox{\bf ]|} \sim K + J \mbox{ } , 
\eeq
the key bracket being the last one by which the boosts are not a subalgebra. 
Thomas precession then refers to the rotation arising thus from a combination of boosts.

\mbox{ } 

\ni IV)  `{\it Two-way integrability}'. If 1) and 4) are nonzero, 
neither $\Frk$ nor $\Frh$ are subalgebras, due to their imposing integrabilities on each other.
I denote this by $\Frk \, \mbox{\textcircled{$\leftrightarrow$}} \, \Frh$.  
In this case, any wishes for $\Frk$ to play a significant role by itself are almost certainly dashed by the actual mathematics of the algebraic structure in question.

\mbox{ } 

\ni III) and IV) have much more diversity of structure than I) and II), and no known systematic means of approaching the corresponding Representation Theory.  

\mbox{ } 

\ni Finally note that Sec's coarse-level treatment of Lie groups readily extends to Lie superalgebras as well \cite{AMech}.

\subsection{Classical constraint algebraic structures}\label{CAS}

One enters the set of constraints in one's possession into the brackets in use to form a constraint algebraic structure.
This may enlarge one's set of constraints, or cause of one to adopt a distinct bracket. 
If inconsistency is evaded, the eventual output is an algebraic structure for all of a theory's constraints, symbolically

\ni \beq
\mbox{\bf |[} \scC_{\sfF} \mbox{\bf ,} \, \scC_{\sfF^{\prime}}\mbox{\bf ]|}_{\sf\si\sn\sa\sll} = {C^{\sfF^{\prime\prime}}}_{\sfF\sfF^{\prime}}\scC_{\sfF^{\prime\prime}} \mbox{ } .  
\eeq
In some cases, the ${C^{\sfF^{\prime\prime}}}_{\sfF\sfF^{\prime}}$ are a Lie algebra's   structure constants, 
whereas in other cases -- in particular for GR,               they are a Lie algebroid's structure functions.  
See Appendix \ref{Algebroids} for a bit more about algebroids.
For now, we note that the preceding Sec's split classification extends to algebroids as well. 
All of the above carries over to supersymmetric versions as well.

\mbox{ }  

\ni Examples of II) Relational mechanics provide a number of examples of semi-direct product constraint algebras. 
There are versions with constraints in correspondence with each of the following:  
the Euclidean group                                               $Eucl(d) = Tr(d) \rtimes Rot(d)$, 
the similarity group                                              $Sim(d)  = Tr(d) \rtimes \{ Rot(d) \times Dil \}$,
the equiareal, (or equivoluminal or higher-$d$ counterpart) group $Equi(d) = Tr(d) \rtimes SL(d, \mathbb{R})$, and 
the affine group                                                  $Aff(d)  = Tr(d) \rtimes GL(d, \mathbb{R})$.
$Tr(d)$ correspond to the zero total            momentum constraint $\underline{\cal P} = \sum_I \underline{P}_I = 0$, 
$Rot(d)$           to the zero total angular    momentum constraint $\underline{\cal L} = \sum_I \underline{Q}^I  \cr  \underline{P}_I$ (or arbitrary-$d$ generalization thereof), and
$Dil$              to the zero total dilational momentum constraint ${\cal D}           = \sum_I \underline{Q}^I \cdot \underline{P}_I$. 
[We are here using $I$ as an index running over the particle labels.]
$SL(d, \mathbb{R})$ corresponds to $\underline{\cal L}$ alongside zero total shear                                                momentum constraints 
                                                              and zero total `Procrustes stretch' ($d$-volume preserving stretch) momentum constraints \cite{AMech}, whereas
$GL(d, \mathbb{R})$ corresponds to all of these alongside ${\cal D}$.

\mbox{ } 

\ni In each case, a suitably compatible form of quadratic energy constraint ${\cal E}$ can be adjoined, but the preceding structures persist as subalgebras of first-class linear constraints.
I.e. 
\beq
\mbox{\bf \{} {\cal F}\ml\mi\mn \mbox{\bf ,} \, {\cal F}\ml\mi\mn \mbox{\bf \}} \sim {\cal F}\ml\mi\mn \mbox{ } , \mbox{ } \mbox{ } 
\eeq
\beq
\mbox{\bf \{} {\cal F}\ml\mi\mn \mbox{\bf ,} \, {\cal E} \mbox{\bf \}} \sim 0 \mbox{ } , \mbox{ } \mbox{ } 
\eeq
\beq
\mbox{\bf \{} {\cal E} \mbox{\bf ,} \, {\cal E} \mbox{\bf \}} \sim 0 \mbox{ } . 
\eeq
The first bracket indicates a closed subalgebra,
the second that ${\cal E}$ is a good ${\cal F}\ml\mi\mn$ object (a scalar), 
and the third indicates that ${\cal E}$ forms its own subalgebra rather than implying any of the ${\cal F}\ml\mi\mn$ as integrabilities.  
Thus adjunction of ${\cal E}$ is a first example of I)

\ni Further examples of I) include a further relational mechanics -- 
a theory \cite{FileR} with $Rot(d) \times Dil$ invariance which models the absense of absolute axes of rotation and absolute scale whilst retaining an absolute origin -- 
provides an example of direct product constraint algebra of $\underline{\cal L}$ and ${\cal D}$.
Of course, the Standard Model also involves a direct product Lie group $SU(3) \times SU(2) \times U(1)$, so $\times$ is well-exemplified within conventional Gauge Theory. 

\mbox{ }

\ni Examples of III) at the level of constraint algebraic structures include (as outlined in \cite{AMech}) the relational mechanics that are based on each of 
the conformal group $Conf(d)$ and on all of the supersymmetric counterparts of the groups in Example II).  
The key bracket here is an integrability of the form 

\ni\beq
\mbox{\bf |[}  {\cal K}  \mbox{\bf ,} \,  {\cal P}  \mbox{\bf ]|} \sim {\cal M} + {\cal D} \mbox{ } ,
\label{K-P}
\eeq 
by which the conformal algebra is $({\cal P}, {\cal K}) \mbox{\textcircled{$\rightarrow$}} ({\cal M}, {\cal D})$ Thomas.
I.e. a translation and a special conformal transformation compose to give both a rotation (`conformal precession') {\sl and} an overall expansion.

\mbox{ }

\ni Then another salient Example of III) is the Dirac algebroid formed by the constraints of GR \cite{Dirac51, Dirac, Tei73, BojoBook}, schematically 
\beq
\mbox{\bf \{} {\cal M}_i \mbox{\bf ,} \, {\cal M}_i \mbox{\bf \}} \sim {\cal M}_i \mbox{ } , \mbox{ } \mbox{ } 
\eeq
\beq
\mbox{\bf \{} {\cal M}_i \mbox{\bf ,} \, {\cal H} \mbox{\bf \}} \sim {\cal H} \mbox{ } , \mbox{ } \mbox{ } 
\eeq
\beq
\mbox{\bf \{} {\cal H} \mbox{\bf ,} \, {\cal H} \mbox{\bf \}} \sim {\cal M} \mbox{ } . 
\label{Dirac-Algebroid}
\eeq
It is the last of these brackets which contains structure functions, with Teitelboim pointing to the interpretational complications of this \cite{Tei73}.
Moreover, the physics of the problem in hand requires this enlargement and complication of algebraic structure, 
to model the myriad of possible foliations of spacetime by spatial hypersurfaces.
Also note that the first of these brackets shows that the momentum constraint ${\cal M}_i$ closes by itself al a subalgebraic structure; 
these are first-class linear constraints corresponding to $Diff(\bupSigma)$.  
It is, moreover a bona fide (if infinite-$d$) Lie algebra, of the spatial diffeomorphisms $Diff(\bupSigma)$ on the corresponding topological manifold $\bupSigma$.
Finally, the second of these brackets means that ${\cal H}$ is a good $Diff(\bupSigma)$ object: a scalar density.  

\mbox{ }

\ni Example of IV) The constraint superalgebroid of Supergravity.  
At the level of algebraic structures, each of supersymmetry and split space-time GR can be separately envisaged as Thomas-type effects.  
Upon considering both at once, these integrabilities furthermore go in opposite directions, so the more complicated `two-way' integrability case arises. 
The schematic form of the key new relation for Supergravity is\footnote{One needs to generalize the Poisson brackets 
to accommodate mixtures of bosonic and fermionic species \cite{Casalbuoni}.
Also the sugra, unlike the unconstrained fermion use, involves a Dirac bracket \cite{Dirac, HT92} here rather than a Poisson bracket.}
\beq
\mbox{\bf \{} {\cal S} \mbox{\bf ,} \, {\cal S} \mbox{\bf \}} \sim {\cal H} \mbox{ } ( \mbox{ } + {\cal M} + {\cal J} \mbox{ } ) \mbox{ } . 
\label{S-S->H}
\eeq
The integrability pointing in the opposite direction remains of the schematic GR form (\ref{Dirac-Algebroid}).  
More details of this superalgebroid can be found at e.g. \cite{DEath, VM10}.

\subsection{Lie algebroids}\label{Algebroids}

\ni {\it Lie algebroids} are defined as follows \cite{Algebroid1, Algebroid2}. 

\mbox{ }

\ni 1) Consider a smooth manifold $\bFrM$, and define a vector bundle $\bFrj$ over this. 

\ni 2) Then define a Lie algebra structure on the corresponding space of sections of $\bFrj$: $\FrS\mbox{ec}(\bFrj)$.

\ni 3) The {\it anchor map} is a bundle map $A: \bFrj \rightarrow \FrT(\bFrM)$ such that 

\ni i)  $A: \FrS\mbox{ec}(\bFrM) \rightarrow \mbox{Vec}(\bFrM)$ (of vectors) is a Lie algebra homomorphism corresponding to the commutator Lie bracket.

\ni ii) For $f \in {\cal C}^{\infty}(\bFrM)$, $\Gamma_1, \Gamma_2 \in Sec(\bFrM)$ the derivation rule 
$\mbox{\bf |[}\Gamma_1 \mbox{\bf ,}\, f \Gamma_2 \mbox{\bf ]|} = f \mbox{\bf |[}\Gamma_1 \mbox{\bf ,}\, \Gamma_2 \mbox{\bf ]|} + (A(\Gamma_1)f)\Gamma_2$ holds. 

\mbox{ } 

\ni The algebroid--groupoid interrelation is now more complicated than its algebra--group counterpart.

\mbox{ } 

\ni Example 1) In the case over just a one-point space, one returns to the standard Lie algebra.  

\mbox{ } 

\ni Example 2) Tangent bundles, tie to foliations and thus to specific case below. 
Here the identity map of $\FrT(\bFrM)$ is the anchor map, and the reps are vector bundles over $\bFrM$ with flat connections.  

\mbox{ } 

\ni Example 3) Lie algebroids arising in the symplectic context are covered in particular in \cite{Algebroid1}.

\mbox{ } 

\ni Example 4) The Dirac alias deformation algebroid \cite{HKT}, as per Appendix \ref{CAS}.

\mbox{ } 

\ni Finally note that a {\it representation in the context of algebroids} \cite{GS08} consists of two parts. 

\mbox{ }

\ni 1) a vector bundle $\bFrj$ over $\bFrM$. 

\ni 2) A $\mathbb{R}$-bilinear map $\FrS\mbox{ec}(\bFrM) \times \FrS\mbox{ec}(\bFrj) \rightarrow \FrS\mbox{ec}(\bFrM): a \otimes s \mapsto D_a s$, 
for a suitable notion of derivative $D_a$ \cite{ELW}.  

\mbox{ }  

\ni The latter is such that for any $a, b \in \FrS\mbox{ec}(\bFrM)$, $s \in \FrS\mbox{ec}(\bFrj)$ and $f \in {\cal C}^{\infty}(\bFrM)$, 
$D_{fa}s = f D_a s$, $D_a\{fs\} = f D_a s + \{\rho(a)f\}s$ and $D_a\{D_b s\} - D_b\{D_a s\} = D_{\mbox{\bf \scriptsize |[}a \mbox{\bf \scriptsize ,} b \mbox{\bf \scriptsize ]|}   } s$.  

\mbox{ }

\ni See e.g. \cite{GS08, Broid-Rep-2} for some Representation Theory methods which extend as far as this case.

\mbox{ } 

\ni Superalgebroids are treated in outline in e.g. \cite{Superalgebroids}; 
the above outline readily continues to carry over upon replacing `Lie algebra' by `Lie superalgebra'; 
Supergravity's constraint superalgebroid is probably Theoretical Physics' most salient example of superalgebroid.

\subsection{Associated algebras}\label{Associated}

`Associated algebras' is meant in this Article in the following sense. 

for $T_{\sfp}$ the generators of an algebraic brackets structure $\Frg$ with bracket operation $\mbox{\bf |[} \mbox{ } \mbox{\bf ,} \, \mbox{ } \mbox{\bf ]|}$, 

\ni \beq
\mbox{\bf |[}T_{\sfp}\mbox{\bf ,} \, V_{\sfV}\mbox{\bf ]|} = 0 
\label{Assoc-Type}
\eeq
produces an associated algebra with respect to the same bracket operation. 

\mbox{ }

\ni Example 1) The case in which the $V_{\sfV}$ themselves are formed from the generators themselves is well-known.
In this case, the association amounts to producing the centre $Z(\Fru)$ of the {\it universal enveloping algebra} $\Fru$ of $\Frg$. 
This is named thus due to its encapsulation of features universal to all representations of the $\Frg$ in question. 
In this case, the resulting $V_{\sfV}$ are known as {\it Casimirs} \cite{AMP2, BojoBook}. 
These play a prominent role in Representation Theory, with $SU(2)$'s total angular momentum operator $J^2$ being the best-known such.
In this example of association procedure, each given Lie algebra leads to a corresponding algebra of Casimirs.  

\mbox{ } 

\ni Example 2) is the Article's main example of observables algebraic structures associated with constraint algebraic structures.

\subsection{Classical observables lemmas}\label{Observables}

\ni {\bf Lemma 1}. Notions of observables can only be meaningfully associated with closed constraint algebraic (sub)structures \cite{ABeables}.  

\ni \underline{Proof}. 
Suppose $O_{\sfO}$ commutes solely with a set of ${\cal C}_{\sfC}$ which is not closed: 
it does not include some of the $\mbox{\bf |[} {\cal C}_{\sfC}\mbox{\bf ,} \, {\cal C}_{\sfC^{\prime}} \mbox{\bf ]|}$.
However, the Jacobi identity with one $B$ and two ${\cal C}$ as entries and making two uses of (\ref{Basic-Obs}) gives

\ni\beq
\mbox{\bf |[}O_{\sfO}\mbox{\bf ,} \,  \mbox{\bf |[}{\cal C}_{\sfF} \mbox{\bf ,} \,  {\cal C}_{\sfC^{\prime}}  \mbox{\bf ]|} \, \mbox{\bf ]|} = - 
\mbox{\bf |[}{\cal C}_{\sfF}\mbox{\bf ,} \,  \mbox{\bf |[}{\cal C}_{\sfF^{\prime}}\mbox{\bf , } O_{\sfO}      \mbox{\bf ]|} \, \mbox{\bf ]|}   - 
\mbox{\bf |[}{\cal C}_{\sfF^{\prime}}\mbox{\bf ,} \,  \mbox{\bf |[}O_{\sfO}\mbox{\bf , } {\cal C}_{\sfF}      \mbox{\bf ]|} \, \mbox{\bf ]|} \approx 0 \mbox{ }  :
\label{CCB} 
\eeq
which is a contradiction.  
Thus such a $\mbox{\bf |[}C_{\sfC} \mbox{\bf ,} \, C_{\sfC^{\prime}} \mbox{\bf ]|}$ in fact has to be included among the quantities $O_{\sfO}$ commutes with. $\Box$

\mbox{ }

\ni {\bf Lemma 2}. The $O_{\sfO}$ themselves close under whichever $\mbox{\bf |[} \mbox{ } \mbox{\bf ,} \, \mbox{ } \mbox{\bf ]|}$ is used to define them.  

\ni \underline{Proof} Take the Jacobi identity with two $B$ and one ${\cal C}$ as entries

\ni\beq
\mbox{\bf |[}{\cal C}_{\sfF}\mbox{\bf ,} \,  \mbox{\bf |[}O_{\sfO} \mbox{\bf ,} \,  O_{\sfO^{\prime}}  \mbox{\bf ]|} \, \mbox{\bf ]|} = - 
\mbox{\bf |[}O_{\sfO}\mbox{\bf ,} \,   \mbox{\bf |[}O_{\sfO^{\prime}}\mbox{\bf , } {\cal C}_{\sfF}      \mbox{\bf ]|} \, \mbox{\bf ]|}   - 
\mbox{\bf |[}O_{\sfO^{\prime}}\mbox{\bf ,} \,  \mbox{\bf |[}{\cal C}_{\sfF}\mbox{\bf , } O_{\sfO}      \mbox{\bf ]|} \, \mbox{\bf ]|} \mbox{ } `=' 0 \mbox{ } ,
\label{BBC}
\eeq
and make two uses of (\ref{Basic-Obs}), one obtains that $\mbox{\bf |[}O_{\sfO}\mbox{\bf ,} \,  O_{\sfO^{\prime}}\mbox{\bf ]|}$ obeys (\ref{Basic-Obs}) as well. $\Box$ 

\mbox{ }

\ni Note moreover that observables algebraic structure $\bFrb$ can be an algebra

\ni\beq
\mbox{\bf |[} O_{\sfO} \mbox{\bf ,} \,  O_{\sfO^{\prime}} \mbox{\bf ]|} = {C^{\sfB^{\prime\prime}}}_{\sfB\sfB^{\prime}}\scO_{\sfO^{\prime\prime}} \mbox{ } , 
\label{observables-Algebra}
\eeq
or an algebroid 

\ni\beq
\mbox{\bf |[} O_{\sfO} \mbox{\bf ,} \,  O_{\sfO^{\prime}} \mbox{\bf ]|} = {C^{\sfB^{\prime\prime}}}_{\sfB\sfB^{\prime}}(\biP, \biQ)\scO_{\sfO^{\prime\prime}} \mbox{ } . 
\label{observables-Algebroid}
\eeq

\subsection{Lattices in outline}\label{Lattices}

\ni A {\it poset} ({\it partially ordered set}) is a set $\FrX$ alongside an ordering relation $\preceq$ which is reflexive, antisymmetric and transitive.
This is less stringent than the more familiar total ordering, for which every two elements have to be related; this permits a linear ranking representation, by which 
total orderings are also known as {\it chains}.  
Posets' elements need not all be related, by which posets in general contain multiple branching chains.  
This is representable using a {\it Hasse diagram}; all the figures in the current paper are based on Hasse diagrams.

\mbox{ } 

\ni A {\it lattice} $\FrL$ is a poset in which each pair of elements has a join (alias least upper bound) and a meet (alias greatest lower bound).

\mbox{ }

\ni An element 1 of $\FrL$ is a {\it unit}         if $\forall \, l \,  \in \,  \FrL$, $l \preceq 1$, 
and an element 0 of $\FrL$ is a {\it null element} if $\forall \, l \,  \in \,  \FrL$, $O \preceq l$. 
A lattice possessing these is called a {\it bounded lattice}; this applies to all examples in the current Article.

\mbox{ } 

\ni A {\it poset morphism} is an order-preserving map between posets, and a {\it lattice morphism} is an order-, join- and meet-preserving map between lattices.  

\mbox{ }

\ni Subset inclusion $\subseteq$ is one common realization of $\preceq$. 
This carries over to the case of sets with further structures, in which case one is dealing with subspace inclusion.  
In this setting, intersection of two subspaces plays the role of meet and smallest subspace containing a pair of spaces that of their join.  
E.g. for subgroups of a group $\lFrg$, the intersection of two subgroups is also a subgroup, and the join of two subgroups is defined as the subgroup generated by their union.
0 = id: the trivial group, and 1 = $\lFrg$ itself.
Constraint subalgebraic structures of an algebraic structure $\FrC$ follow suit in this regard, with 0 = id: the trivial algebra, and 1 = $\FrC$ itself.  
On the other hand, for observables subalgebraic structures of an observables algebraic structure $\FrO$ associated a with constraint algebraic structure $\FrC$, 
each constraint present in the latter induces an extra relation in the former.
Hence the association map $Assoc$ is an order-reversing lattice morphism $\FrC \longrightarrow \FrO$.  
This can be recast as order-preserving by use of $\supseteq$ in place of $\subseteq$.
$\FrO$ still possesses a 0 and a 1, with, by the order-reversing, $1_{\FrO} = Assoc(0_{\FrC}) = Assoc(id)$ the space $\FrU$ of unconstrained observables $U_{\sfU}$ and 
                                                                  $0_{\FrO} = Assoc(1_{\FrC}) = Assoc(\FrC)$ the space $\FrD$ maximally constrained --i.e. Dirac -- observables $D_{\sfD}$.
Hence the lattice of observables is also a bounded lattice.

\mbox{ }

\ni Note that $Assoc$ maps all constraint subalgebraic structures to observables algebraic substructures, since one can always pose the problem of what functions 
commute with a given constraint subalgebraic structure.
Adding constraints can never increase the freedom in the observables algebraic structure, but on occasion it may have no effect if the constraint being added imposes 
no further restriction on the allowed functional form of the observables.
E.g. an angle quantity is already a ratio and a combination of dot products.
I.e. $Assoc$ is capable of being many to one, as occurs in the example of conformal relational mechanics.
In such a case, $Assoc$ is a coarse-graining of the ordering information in moving between the two lattices; 
also note that in all the other examples considered in this article, $Assoc$ is 1 : 1 and fully preserving of the ordering information.  

\mbox{ }

\ni See \cite{IshamBook} for a bit more about lattices with basic and theoretical physics examples, and \cite{Cohn2} for somewhat more theory.  
With many conceptually and technically interesting properties of lattices having been worked out, it will be very interesting to see which of these apply to lattices of A-observables.

\end{appendices}


\end{document}